\documentclass[conference]{IEEEtran}
\IEEEoverridecommandlockouts
\usepackage{cite}
\usepackage{amsmath,amssymb,amsfonts}
\usepackage{graphicx}
\graphicspath{{Figures/}} %Setting the graphicspath
\usepackage[caption=false,font=footnotesize]{subfig}
\usepackage{floatrow}
\usepackage{float}
\floatstyle{plaintop}
\restylefloat{table}
\usepackage{multirow}
\usepackage{mathtools}
\usepackage{siunitx}
\usepackage{epstopdf}
\usepackage{xcolor} 
\definecolor{darkgreen}{rgb}{0.0, 0.5, 0.0}
\definecolor{tableblue}{rgb}{0.2, 0.2, 0.8}
\usepackage{textcomp}
\usepackage{booktabs}
\usepackage[table]{xcolor}
\usepackage{xcolor}
\usepackage{multirow}
\usepackage[font=footnotesize]{caption}
\usepackage{color,soul}
\usepackage{setspace}
\newcommand{\name}{HyMGP} 

\usepackage{stfloats}% <-- added
\usepackage[strings]{underscore}
\usepackage[left=1.62cm,right=1.62cm,top=0.7in]{geometry}
\usepackage{float}
\usepackage[draft]{hyperref}
\begin{document}
\title{Optimizing Hybrid Wind-Solar Systems: A Techno-Economic Perspective for Remote Microgrids}
\title{MILP-Based Planning of Islanded Hybrid Microgrids: A Comparative Techno-Economic Analysis with HOMER Pro}
\title{Techno-Economic Optimization of Islanded Microgrids Using a Customized MILP-Based Tool}
\title{MILP-based Planning of Islanded Microgrids: Performance and Economic Evaluation}
%\title{MILP-Based Planning of Islanded Microgrids: Optimization, Performance, and Cost Analysis}
\title{HyMGP: A Customized MILP-Based Tool for Techno-Economic Planning of Islanded Microgrids}
\IEEEaftertitletext{\vspace{-2\baselineskip}}
\author{
    \IEEEauthorblockN{
        Andres Intriago\IEEEauthorrefmark{1}, Rongxing Hu\IEEEauthorrefmark{1}, Nabil Mohammed\IEEEauthorrefmark{1}, S. Gokul Krishnan\IEEEauthorrefmark{1},\\ Konstantinos Kotsovos\IEEEauthorrefmark{2}, Issam Gereige\IEEEauthorrefmark{2}, Nesren Attiah\IEEEauthorrefmark{2}, Ali Basaheeh\IEEEauthorrefmark{2}, Sarah Aqeel\IEEEauthorrefmark{2},\\ Hamad A. Saiari\IEEEauthorrefmark{2}, Shehab Ahmed\IEEEauthorrefmark{1},  and Charalambos Konstantinou\IEEEauthorrefmark{1}
    }
    
    \IEEEauthorblockA{\IEEEauthorrefmark{1}CEMSE Division, King Abdullah University of Science and Technology (KAUST)}
    
    \IEEEauthorblockA{\IEEEauthorrefmark{2}Saudi Aramco R\&D Center, KAUST}

}

\markboth{}%
{Shell \MakeLowercase{\textit{et al.}}: Control Frame Work for DC Microgrids}
\markboth{}%
{Shell \MakeLowercase{\textit{et al.}}: Control Frame Work for DC Microgrids}
%\IEEEaftertitletext{\vspace{-1\baselineskip}}
\maketitle

\begin{abstract}

This paper presents a customized microgrid planning algorithm and tool, \name, for remote sites in arid regions, which is formulated as a Mixed Integer Linear Programming (MILP) problem. \name\ is compared with HOMER Pro to evaluate its performance in optimizing the sizing of microgrid components, including photovoltaic panels (PVs), vertical axis wind turbines (VAWTs), and battery energy storage systems (BESS), for remote and off-grid applications. The study focuses on a standalone microgrid in the Saudi Arabia, considering high solar irradiance, limited wind availability, and a constant load profile composed of continuous cathodic protection and daytime cooling. In the simulation environment, comparisons with HOMER solutions demonstrate the advantages of \name, which provides optimal and more flexible solutions by allowing user-defined component specifications and strictly enforcing all constraints. Further analysis shows that incorporating wind turbines reduces the Net Present Cost (NPC) by decreasing the required PV and battery capacities. Increasing battery autonomy leads to a higher NPC in both PV-only and hybrid systems due to the need for larger storage. Finally, lithium iron phosphate (Li-ion LFP) batteries are found to be more cost effective than lead acid, offering lower NPCs due to their longer lifespan, deeper discharge capability, and fewer replacement cycles.

\end{abstract}


\IEEEpeerreviewmaketitle

%\vspace{-2mm}
\section{Introduction}\label{s:introduction}
%\vspace{-1mm}
Over the years, electric power systems (EPS) have shifted from a centralized, unidirectional generation and transmission framework to a more distributed architecture that incorporates both conventional generation and distributed energy resources (DERs), including distributed generation (DGs) such as photovoltaic panels (PVs) and wind turbines (WTs), as well as distributed storage (DS) systems including battery energy storage systems (BESS) \cite{9340265}. 
This evolution is particularly important in remote and off-grid locations, where the absence of grid-connected infrastructure makes it difficult to ensure a reliable and cost-effective power supply. In these cases, microgrids, which are small-scale power systems powered by local DERs, are considered a viable solution \cite{Bidram2013}.
Microgrids operating in islanded mode need to function independently without support from the main grid \cite{Intriago2024}. In this mode, the microgrid is fully responsible for maintaining its own voltage and frequency levels \cite{Guerrero2011}, while continuously supplying power to local loads. Given these requirements, microgrid planning in remote and harsh environments requires a careful balance between reliability, autonomy, and economic performance.

This study focuses on a hybrid microgrid designed for deployment in a remote desert region of Saudi Arabia, where research on microgrid planning is relatively limited. The selected site experiences high solar irradiance levels, low wind availability, and significant temperature fluctuations. The electrical load consists mainly of critical systems, including cathodic protection and cooling modules, which require continuous and stable power throughout the year. To address this, the proposed system integrates PV generation, a vertical-axis wind turbine (VAWT), and a BESS, creating a standalone configuration that could supply energy to the load. To effectively perform microgrid planning, particularly in this remote and off-grid context addressed in this study, dedicated algorithm and software tools are required to simulate system behavior, evaluate cost-effectiveness, and optimize component sizing. 

Table~\ref{tab:software_comparison} indicates several optimization tools employed to perform system planning, including SAM, DER-CAM, MDT, REopt, and HOMER Pro, each one offering specific functionalities. Among the tools listed, HOMER Pro is selected in this study as a benchmark for comparison due to its widespread use and proven capabilities in off-grid hybrid system design. It offers key features such as search space exploration, which identifies optimal system configurations and component sizing strategies; sensitivity analysis, which evaluates the impact of resource variability and load changes on system costs; and hybrid system optimization, which enables coordinated planning of PV arrays, BESS, and WTs. HOMER Pro incorporates financial indicators such as Net Present Cost (NPC) and Levelized Cost of Energy (LCOE), supporting robust techno-economic evaluation. However, its optimization process relies on predefined heuristics, limiting user control over constraints and dispatch strategies, and cannot support uncertainty modeling. HOMER Pro models the behavior of the components using first degree equations, which do not represent the complexity and nonlinear characteristics of PV, BESS, and WTs \cite{Erdinc2012}. 

To address these limitations, customized algorithms have been developed for islanded microgrid planning, taking into account the site resources, such as solar irradiance, and wind speed, the component ratings, and electricity generation costs. One approach involves the use of Mixed Integer Nonlinear Programming (MINLP) to determine the optimal sizes of DER components, including PV panels, BESS, and AC/DC converters \cite{Mohamed2019}. Due to the high computational complexity of such problems, meta-heuristic algorithms, such as Whale Optimization Algorithm (WOA), Particle Swarm Optimization (PSO), and Genetic Algorithm (GA) have also been adopted to effectively compute optimal component sizes \cite{Shezan2023}. These methods enable flexible consideration of system constraints and incorporate demand-side management (DSM) strategies tailored to standalone microgrids \cite{pso_sizing}. Alternatively, Mixed-Integer Linear Programming (MILP) provides a structured method to optimally size microgrid components, such as PVs, WTs, diesel generators, and BESS, while minimizing investment, operational, and maintenance costs \cite{Zelaschi2025}.

The contributions of this paper are threefold. First, we develop a customized MILP-based microgrid planning algorithm tailored for remote sites in the Saudi Arabian desert. Second, we conduct a comparison between the developed algorithm and HOMER Pro, demonstrating the advantages of the former in enabling fully user-defined component specifications, enforcing operational constraints, and obtaining optimal solutions using powerful off-the-shelf solvers. Third, we perform sensitivity analysis on battery autonomy and the type of BESS technologies to provide insights for planning projects in similar remote environments. The proposed tool is well-suited for designing practical standalone microgrids, where a stable and uninterrupted electricity supply is essential to ensure continuous operation of the hybrid system and the connected load.

\begin{table}[!t]
\scriptsize
\centering
\caption{Existing microgrid design tools in literature.}
%\vspace{-1mm}
\label{tab:software_comparison}
\setlength{\tabcolsep}{2pt}
\renewcommand{\arraystretch}{1.15}
\begin{tabular}{|l|p{1.6cm}|p{2cm}|p{2cm}|}
\hline
\textbf{Software} & \textbf{Components} & \textbf{Functionalities} & \textbf{Optimization Algorithm} \\
\hline
SAM \cite{Blair2017} & PVs, WTs, Hydro & Hybrid optimization, techno-economic analysis & Built-in \\
\hline
DER-CAM \cite{Mathur2017} & PVs, WTs, Fossil, Fuel Cells, CHP & DER sizing and dispatch optimization & MILP \\
\hline
MDT \cite{Mathur2017} & PVs, WTs, Fossil & Sizing, reliability assessment & MILP \\
\hline
REopt \cite{Mathur2017} & PVs, WTs, Biomass, CHP & Techno-economic optimization & MILP \\
\hline
HOMER Pro \cite{Mathur2017} & PVs, WTs, BESSs, Diesel, Hydro & Hybrid optimization, economic and sensitivity analysis & Built-in heuristic \\
\hline
HyMGP & PVs, WTs, BESSs, Diesel & Full-featured hybrid optimization & Custom solvers and constraints \\
\hline
\end{tabular}
\vspace{-4mm}
\end{table}
% \fi

%\vspace{-2mm}
\section{Optimal Hybrid Microgrid Formulation}\label{s:simulation}

\subsection{Objective Function}
The goal of the optimal planning algorithm is to determine the design of the microgrid that minimizes the total NPC while satisfying the operating specifications of the components and the requirements of the microgrid system. The total NPC includes capital costs, replacement costs, operation and maintenance (O\&M) costs, and the salvage/residual value of all components. The decision variables are the number of units for each component, denoted as $N_x$. The objective function is as follows: 

\begin{equation}
    f_{\text{total}}^{\text{npc}} = \sum_{x \in \{\text{wt}, \text{pv}, \text{bess}\}} N^x C_{\text{npcUnit}}^x
    \label{eq:obj}
\end{equation}
\begin{equation}
    C_{\text{npcUnit}}^x = C_{\text{capUnit}}^x + C_{\text{repUnit}}^x + C_{\text{OMUnit}}^x - S_{\text{Unit}}^x
    \label{eq:obj_npc}
\end{equation}
\begin{equation}
    C_{\text{repUnit}}^x = \sum_{y=1}^{Y} \frac{C_{\text{repUnit},y}^x}{(1 + \alpha)^y}
    \label{eq:obj_rep}
\end{equation}
\begin{equation}
    C_{\text{OMUnit}}^x = \sum_{y=1}^{Y} \frac{C_{\text{OMUnit},y}^x}{(1 + \alpha)^y}
    \label{eq:obj_OM}
\end{equation}
\begin{equation}
    S_{\text{Unit}}^x = C_{\text{capUnit}}^x \frac{(1 + N_{\text{rep}}^x) l_x - Y}{l_x (1 + \alpha)^Y}
    \label{eq:obj_salv}
\end{equation}
\begin{equation}
    N_{\text{min}}^x \leq N^x \leq N_{\text{max}}^x
    \label{eq:obj_var_bound}
\end{equation}
\begin{equation}
    P_{\text{rate}}^x = N^x P_{\text{unit}}^x, \! \, x \in \{\text{pv}, \text{wt}\}
    \label{eq:obj_rate_pv}
\end{equation}
\begin{equation}
    E_{\text{rate}}^x = N^x E_{\text{unit}}^x, \! \, x \in \{\text{bess}\}
    \label{eq:obj_rate_bess}
\end{equation}
\begin{equation}
    P_{\text{rate}}^x = {E_{\text{rate}}^x}/{h_{\text{full}}},\! \, x \in \{\text{bess}\}
    \label{eq:obj_rate_bess_p}
\end{equation}

\noindent where $C_{\text{npcUnit}}^x$ denotes the NPC of a single component unit, $C_{\text{capUnit}}^x$ represents the initial capital cost, $C_{\text{repUnit}}^x$ denotes the replacement cost, $C_{\text{OMUnit}}^x$ represents the O\&M cost, and $S_{\text{Unit}}^x$ is the salvage value at the end of the project life. $Y$ is the project horizon, $y$ is the year index, $\alpha$ represents the discount rate. $N_{\text{rep}}^x$ is the number of replacements that occur, and $l_x$ is the lifespan of component $x$. $N_{\text{min}}^x$ and $N_{\text{max}}^x$ are the minimum and maximum number of components. $P_{\text{unit}}^x$ and $E_{\text{unit}}^x$ are the sizes of component units, $P_{\text{rate}}^x$ and $E_{\text{rate}}^x$ are the total rating of components, and $h_{\text{full}}$ is the charging hours when the BESS charges at its power rating from 0\% state of charge (SoC) to 100\% SoC.

%\vspace{-2mm}
\subsection{Operational Constraints}
The PV, WT, BESS related operational constraints:
\begin{equation}
     0 \leq P_t^{\text{pv}} \leq P_{\text{rate}}^{\text{pv}} p_t^{\text{pvPred}}
    \label{eq:constr_pv}
\end{equation}
\begin{equation}
    0 \leq P_t^{\text{wt}} \leq P_{\text{rate}}^{\text{wt}} p_t^{\text{wtPred}}
    \label{eq:constr_wt}
\end{equation}
\begin{equation}
    0 \leq P_t^{\text{dch}} \leq U_t^{\text{dch}} P_{\text{rate}}^{\text{bess}}
    \label{eq:constr_bess_dch}
\end{equation}
\begin{equation}
    0 \leq P_t^{\text{ch}} \leq (1 - U_t^{\text{dch}}) P_{\text{rate}}^{\text{bess}}
    \label{eq:constr_bess_ch}
\end{equation}
\begin{equation}
    P_t^{\text{bess}} = P_t^{\text{dch}} - P_t^{\text{ch}}
    \label{eq:constr_bess_p}
\end{equation}
\begin{equation}
    E_t^{\text{bess}} = E_{t-1}^{\text{bess}} + \left( \eta^{\text{bess}} P_t^{\text{ch}} - {P_t^{\text{dch}}}/{\eta^{\text{bess}}} \right) \Delta t
    \label{eq:constr_bess_balan}
\end{equation}
\begin{equation}
    SOC_{\text{min}} E_{\text{rate}}^{\text{bess}} \leq E_t^{\text{bess}} \leq SOC_{\text{max}} E_{\text{rate}}^{\text{bess}}
    \label{eq:constr_bess_soc}
\end{equation}
\noindent where $t$ and $\Delta t$ are the time index and interval, and $T$ is the total time steps. $P_t^{\text{pv}}$ and $P_t^{\text{wt}}$ denote the utilized PV and WT power, $p_t^{\text{pvPred}}$ and $p_t^{\text{wtPred}}$ represent the corresponding predictions for per kW PV and WT, which can be calculated using the predicted irradiance and wind speed. $P_t^{\text{dch}}$ and $P_t^{\text{ch}}$ represent the charging and discharging power of the BESS, $U_t^{\text{dch}}$ is a binary variable that indicates the discharging status. $\eta^{\text{bess}}$ represents the charging and discharging efficiency. $P_t^{\text{bess}}$ and $E_t^{\text{bess}}$ are the BESS power and stored energy. $SOC_{\text{min}}$ and $SOC_{\text{max}}$ denote the min. and max. SOCs of the BESS.

The microgrid needs to ensure load and supply balance and maintain a certain level of system reserve. Since serving the load during extreme conditions, such as no irradiance, is very challenging and requires significantly oversized capacity, users may accept a small amount of unserved load $P_t^{\text{UL}}$ or unmet reserve $P_t^{\text{UR}}$ to achieve an appropriate tradeoff between system cost and reliability. Therefore, the power balance and reserve requirements are relaxed by introducing $P_t^{\text{NS}}$ and $P_t^{\text{NR}}$, which can be set according to the considerations of the users.
\begin{equation}
    P_t^{\text{pv}} + P_t^{\text{wt}} + P_t^{\text{bess}} + P_t^{\text{UL}} \geq P_t^{\text{load}}
    \label{eq:constr_sys_balan}
\end{equation}
\begin{equation}
    P_t^{\text{pvPred}} + P_t^{\text{wtPred}} + P_{\text{rate}}^{\text{bess}} + P_t^{\text{UR}} \geq (1 + k^{\text{res}}) P_t^{\text{load}}
    \label{eq:constr_sys_res}
\end{equation}
\begin{equation}
    P_t^{\text{UL}}, P_t^{\text{UR}} \geq 0
    \label{eq:constr_unmet_p}
\end{equation}
\begin{equation}
    \sum_{t \in T} P_t^{\text{UL}} \leq k^{\text{UL}} \sum_{t \in T} P_t^{\text{load}}
    \label{eq:constr_unmet_load}
\end{equation}
\begin{equation}
    \sum_{t \in T} P_t^{\text{UR}} \leq k^{\text{UR}} \sum_{t \in T} P_t^{\text{load}}
    \label{eq:constr_unmet_res}
\end{equation}
\noindent where $k^{\text{res}}$ denotes the reserve factor. $P_t^{\text{load}}$ is the load. $k^{\text{UL}}$ and $k^{\text{UR}}$ are the factors that limit the maximum unserved total load and unmet total reserve, respectively.

The optimal microgrid planning problem is formulated as a MILP problem and solved using Gurobi in MATLAB on a computer with an Intel i7-12700K processor and 8 GB RAM.

% \vspace{-2mm}
\section{Simulations and Results}\label{s:results}

\begin{figure}[!t]
    \centering
    \includegraphics[width=0.8\columnwidth]{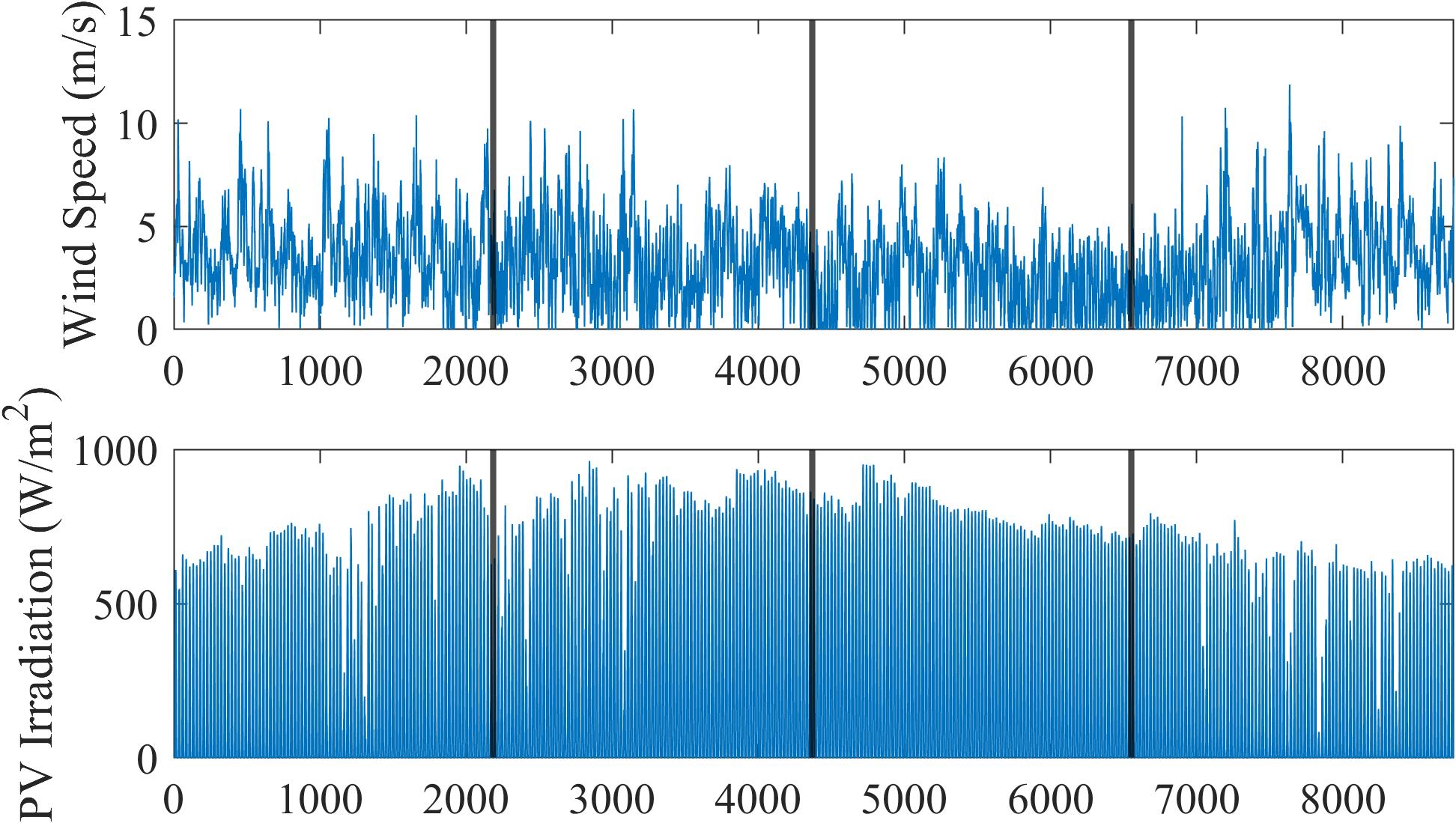}
    %\vspace{-2mm}
    \caption{The hourly wind and irradiance data in 2018 at the remote off-grid location.}
    \label{fig:solar_wind_data}
\end{figure}

The hourly irradiance and wind speed data for an entire year are used for microgrid planning. The data employed in the case study are shown in Fig.~\ref{fig:solar_wind_data}. Component unit costs from our recent projects are provided in Table~\ref{tab:unit_costs_final}. The microgrid is designed considering a 48 V DC bus, and the components such as the PV array and the battery bank, are designed to operate properly at this voltage through series string configurations, a string serves as the component unit considered in the planning process. A VAWT with a rated power of 2 kW is integrated into the hybrid system. This choice is motivated by the wind conditions at the remote site, which has low wind speeds and variable directions. VAWTs are better suited for such environments given their ability to operate efficiently under low-speed and turbulent wind availability \cite{Eftekhari2022}. In the base cases, lead-acid (LA) batteries that can fully charge or discharge in 5 hours are considered. The project lifetime is assumed to be 25 years. The system reserve factor \(k^{\text{res}}\) is set to 15\%, while the sensitivity analysis will be performed on \(k^{\text{UL}}\). The discount rate is 5\%.

\subsection{Simulation Settings in HOMER Pro}
\begin{figure}[t]
    \centering
    \includegraphics[width=0.55\columnwidth]{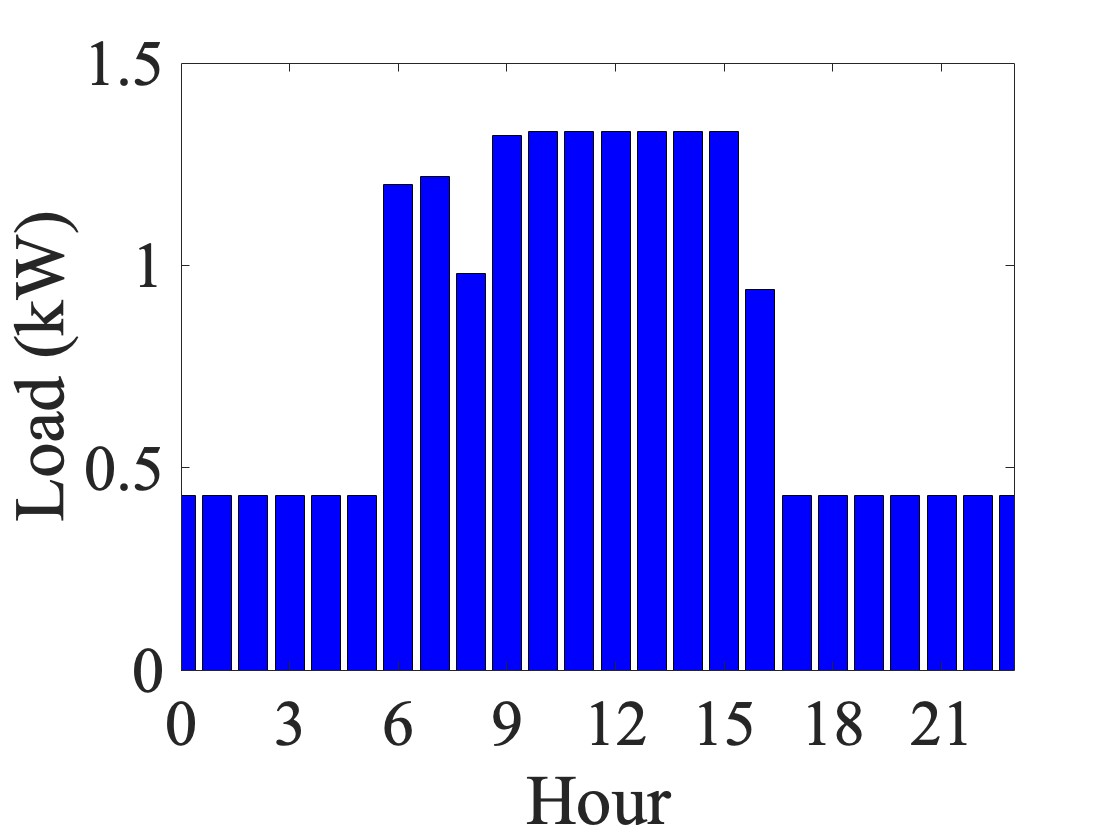}
    %\vspace{-4mm}
    %\vspace{2mm}
    \caption{Daily profile of the considered load.}
    %\vspace{-6mm}
    \label{fig:daily_load_profile}

\end{figure}

The selected microgrid deployment site is located in the Eastern Province of Saudi Arabia. The dataset used in this analysis comprises two key meteorological variables: wind speed and solar irradiance, collected over 2018. The load profile used in this study reflects the energy demands of a remote infrastructure system located in a desert environment. As illustrated in Fig. \ref{fig:daily_load_profile}, a constant load of 430 W corresponds to the cathodic protection system, which operates over the full day. During daylight hours, five cooling modules activate due to high temperatures, increasing the total power demand to a peak of 1.33kW and a total daily energy consumption of 19.01 kWh.
 Fig. \ref{fig:solar} shows the monthly solar radiation levels at the remote site during 2018. The average solar radiation is 5.22 kWh/m\textsuperscript{2}/day. The highest value is recorded in June at 6.990 kWh/m\textsuperscript{2}/day. The lowest irradiance is observed in November, with 3.296 kWh/m\textsuperscript{2}/day. These seasonal variations are consistent with the regional climate in Saudi Arabia, where solar radiation peaks during the summer months and decreases during winter. Fig. \ref{fig:wind} shows the monthly wind speed during 2018. The annual average wind speed is 3.29 m/s. The highest value is recorded in January at 3.982 m/s. The lowest value occurs in September at 2.055 m/s.

\iffalse
\begin{figure}[t]
    \centering
    \includegraphics[width=0.8\columnwidth]{Figures/solar_irradiance.png}
    %\vspace{2mm}
    \caption{Monthly average solar irradiance data for the remote microgrid location.}
    %\vspace{-6mm}
    \label{fig:solar_irradiance}
\end{figure}

\begin{figure}[t]
    \centering
    %\vspace{-3mm}
    \includegraphics[width=0.8\columnwidth]{Figures/wind_speed.png}
    %\vspace{2mm}
    \caption{Monthly average wind speed data for the remote microgrid location.}
    %\vspace{-6mm}
    \label{fig:wind_speed}
\end{figure}

\fi

\iffalse
\begin{figure}[!t]
    \centering
    %\vspace{-3mm}
    \begin{minipage}[b]{0.49\linewidth}
        \centering
        \includegraphics[width=\linewidth]{Figures/solar.jpg}
        \subcaption{Monthly average solar irradiance}
        \label{fig:solar}
    \end{minipage}
    \hfill
    \begin{minipage}[b]{0.49\linewidth}
        \centering
        \includegraphics[width=\linewidth]{Figures/wind.jpg}
        \subcaption{Monthly average wind speed}
        \label{fig:wind}
    \end{minipage}
    \caption{Meteorological conditions at the remote site.}
    %\vspace{-3mm}
    \label{fig:meteo}
\end{figure}
\fi

\begin{figure}[!t]
    \centering
    \subfloat[Monthly average solar irradiance]{%
      \includegraphics[width=0.49\linewidth]{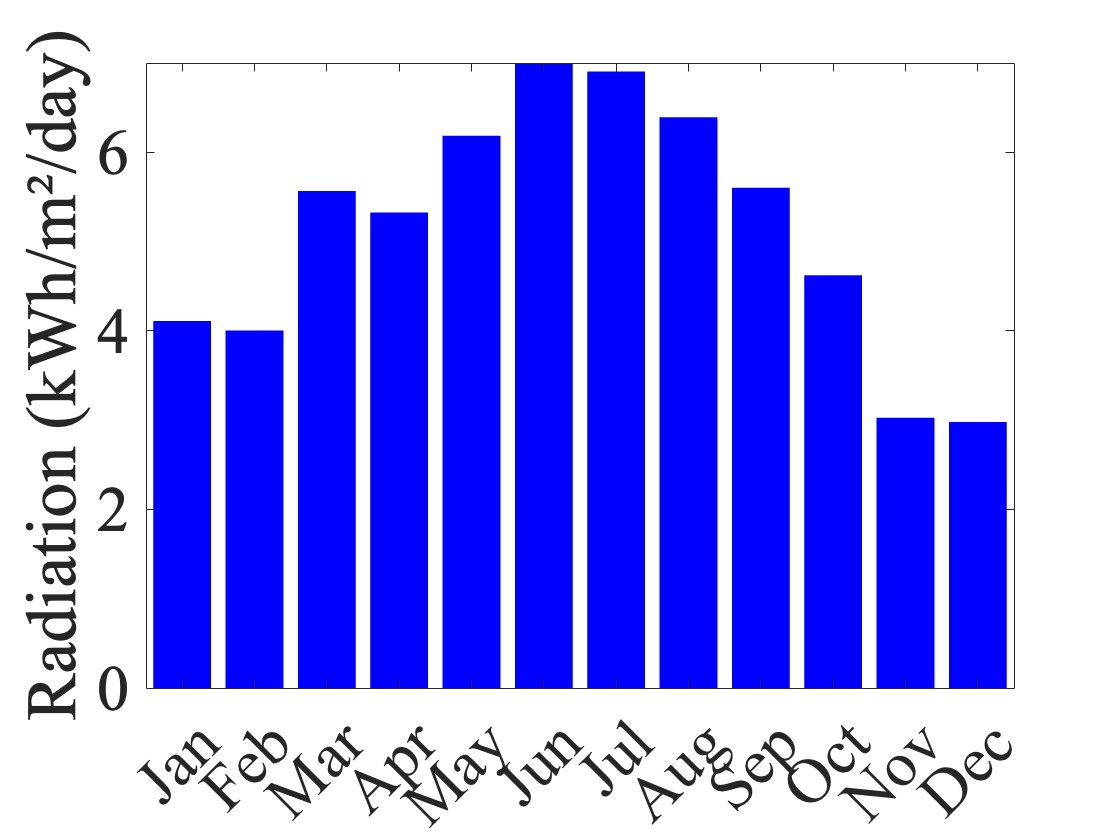}%
      \label{fig:solar}}
    \hfill
    \subfloat[Monthly average wind speed]{%
      \includegraphics[width=0.49\linewidth]{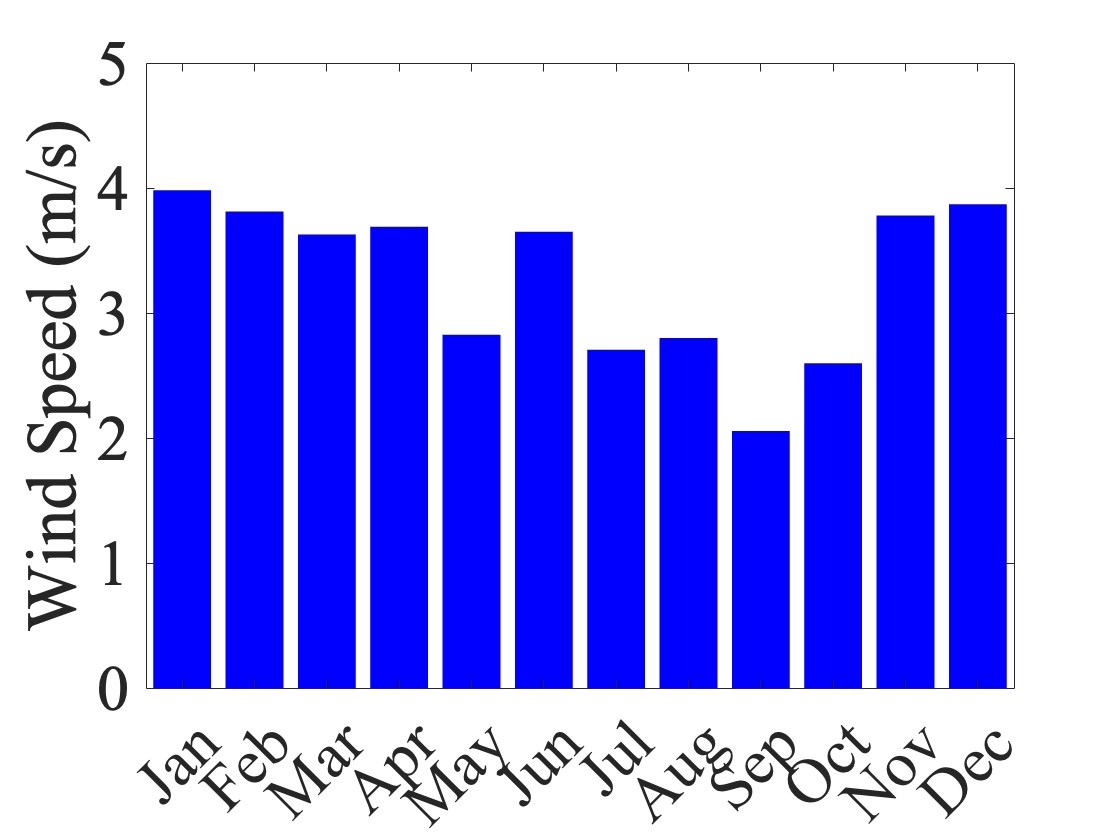}%
      \label{fig:wind}}
    \caption{Meteorological conditions at the remote site.}
    \label{fig:meteo}
%\vspace{-4mm}
\end{figure}

\begin{table}[!t]
    \centering
    \caption{Component unit size and total cost including installation.}
     %\vspace{-3mm}
    \label{tab:unit_costs_final}
    \scriptsize
    \setlength{\tabcolsep}{4pt}
    \renewcommand{\arraystretch}{1.3}
    \begin{tabular}{|l|c|c|c|}
        \hline
        \textbf{Component} & \textbf{Unit Size} & \textbf{Lifespan (year)}& \textbf{Price w/. installation (\$)} \\
        \hline
        PV  & 0.1 kW & 25 & 348.85 \\
        Wind  & 2.0 kW & 25 & 5617.92 \\
        BESS-LA & 9.32 kWh & 4 & 7951.49 \\
        BESS-LFP & 6.24 kWh & 10 & 8619.17 \\
        \hline
    \end{tabular}
\vspace{-4mm}
\end{table}

To perform an accurate economic assessment of the microgrid, this study adopts a per unit capital expenditure (CAPEX) calculation methodology, where this metric represents the total expense of owning and operating one physical unit. The assessment is based on the capital costs in Table \ref{tab:unit_costs_final}, the corresponding operation and maintenance (O\&M) costs, replacement costs, and salvage/residual values. To provide a more detailed and representative cost breakdown, the component capital costs listed include four key parts: the cost of purchasing the equipment, inverter (for PV and wind systems), shelter (applicable to the battery system), and communication and surveillance infrastructure, which applies to the components in the microgrid. In addition, the techno-economic assessment accounts for replacement costs based on component lifespans. We focus on the substitution of physical components, excluding global ancillary expenses such as communication, security, and surveillance. The O\&M costs associated with components throughout the project lifetime are computed as 1.5\% of the corresponding capital costs. Based on the capital cost, O\&M, replacement cost, salvage value, and the specifications of the selected components, HOMER Pro provides planning solutions that minimize the NPC and determine the appropriate sizing of each component.

\begin{figure}[!t]
    \centering
    \includegraphics[width=0.55\linewidth]{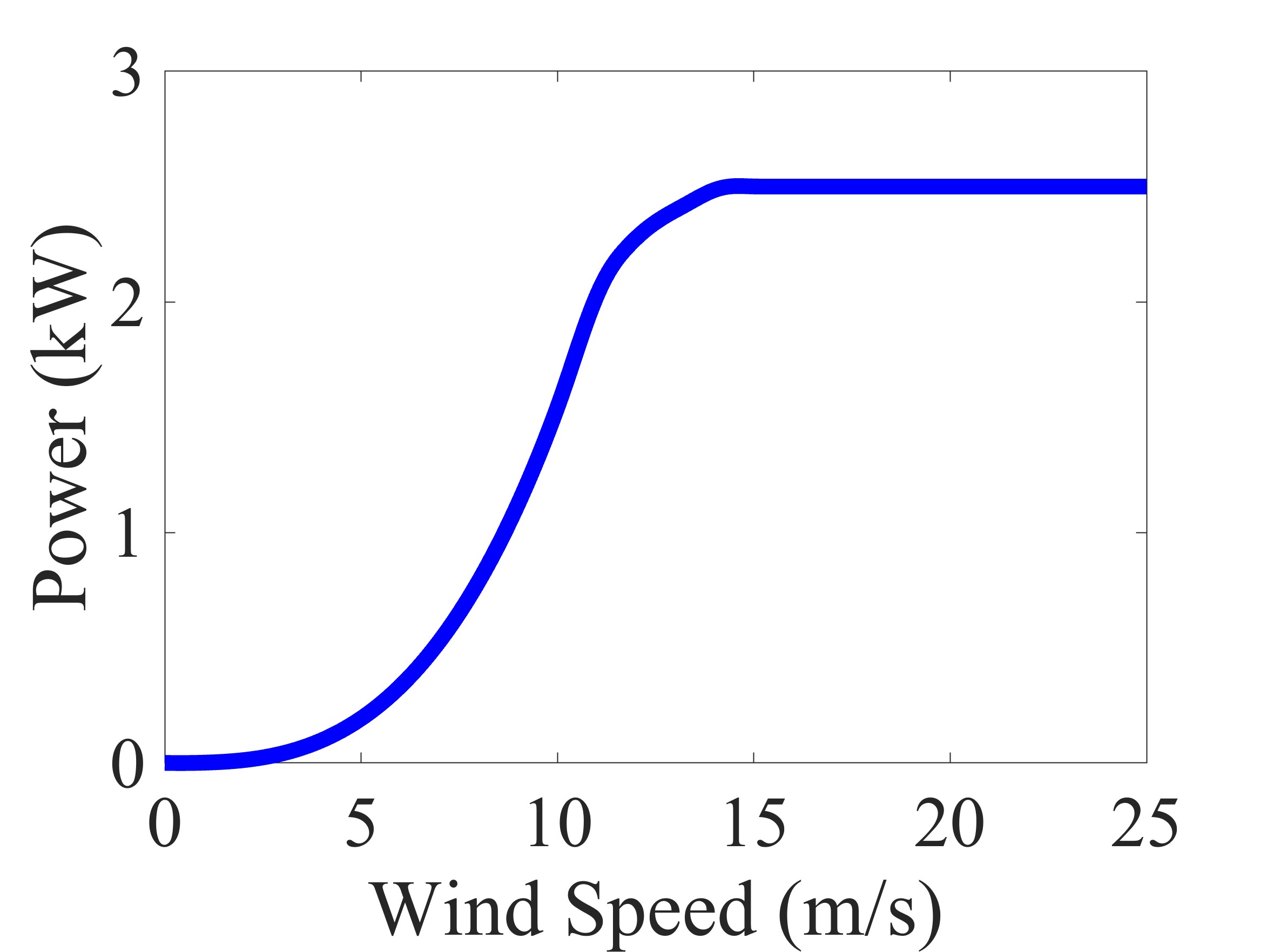}
    \caption{Power curve of the GV-2KW VAWT.}
    \label{fig:wt_power_curve}
    \vspace{-3mm}
\end{figure}

Fig. \ref{fig:wt_power_curve} illustrates the GV-2kW power curve of the VAWT, showing the nonlinear relationship between wind speed and power generation. The turbine produce power at a cut-in speed of 2.8 m/s, reaches its rated output at approximately 11 m/s, and maintains a constant maximum power of 2.5 kW up to its survival wind speed of 25 m/s. However, HOMER Pro does not have a VAWT with a 2 kW rating. Therefore, we use the 1 kW Aeolos V model instead, defining the same characteristics as the GV-2kW and restricting the search space to multiples of 2. The PV system used in this study is modeled as a generic flat-plate PV panel with a rated capacity of 0.1 kW. The module is provided by a generic manufacturer, assumes a derating factor of 100\%, and has an expected operational life of 25 years. Two battery technologies are considered: the initially selected BAE Secura lead-acid battery and the lithium iron phosphate (Li-Ion LFP) battery. The lead-acid unit operates at 48 V, consisting of 4 battery cells in a series string. Each cell has a nominal capacity of 194 Ah and 2.33 kWh of energy. The lead-acid unit is a 5-hour battery and is assumed to be replaced every 4 years. In contrast, a Li-ion LFP battery cell operates at 48 V and serves as a unit, with a nominal capacity of 130 Ah and 6.24 kWh of energy. It has a longer replacement interval of 10 years and is a 1-hour battery with fast charging and discharging capability. For simulation purposes, the lead-acid system is configured with a minimum SoC of 40\%, while the LFP battery is set at 20\%. The comparative performance of both technologies is analyzed in the next subsection.

\subsection{Solution Comparison of HOMER Pro and \name}
Please note that HOMER Pro provides fast solutions that may not ensure all system settings strictly, one instance we observed is the unmet load. Therefore, we simulated cases with 0\% and 0.05\% unmet load (or expected energy not served, EENS) using the developed MILP algorithm. Table~\ref{tab:sol_opti_lead}  presents the detailed solutions obtained from HOMER Pro and the MILP algorithm: 

a) Although the total capacity of renewables is increased, adding WT contributes to lower NPC by decreasing the required capacity of BESS, the reductions are 18\%, 28\%, and 23\% for the three cases, respectively. 

b) Significant cost reduction can be achieved when a small amount of unmet load or load shedding is allowed. Using the developed algorithm, allowing 0.05\% unmet load over the year results in cost reductions of 5.4\% and 10.8\% for the PV-only and PV+WT hybrid cases, respectively. This is because there are several days (as shown in Fig.~\ref{fig:solar_wind_data}) with very low renewable generation, and supplying those days would otherwise require significantly more generation and storage capacity.

c) The developed method can obtain more optimal solutions while ensuring all constraints are strictly satisfied. Although unmet load is set to zero in HOMER Pro, its solutions exhibit 0.06\% and 0.04\% unmet load for the PV and PV+WT hybrid cases, respectively, whereas the developed algorithm fully satisfies the constraints. Furthermore, with similar levels of unmet load (0.04\% in HOMER Pro vs. 0.05\% in our method), the developed algorithm achieves an optimal solution with 10\% lower NPC for the PV+WT case compared to HOMER. Although the developed algorithm requires 130 seconds to compute the solution, compared to less than 60 seconds for HOMER Pro, computation time is acceptable for planning problems.

\begin{table}[!t]
    \caption{Optimal planning results (Lead Acid).}
    %\vspace{-1mm}
     % \vspace{-3mm}
    \label{tab:sol_opti_lead}
    \centering
    \scriptsize
    \setlength{\tabcolsep}{3pt} % Reduce column padding
    \renewcommand{\arraystretch}{1.2} % Adjust row height for better vertical centering
    \begin{tabular}{|c|c|c|c|c|c|c|}
        \hline
        \textbf{Method} & \multicolumn{2}{c|}{\textbf{HOMER Pro}} & \multicolumn{4}{c|}{\textbf{Developed}} \\
        \hline
        \textbf{Unmet} & \multicolumn{2}{c|}{\textbf{0}} & \multicolumn{2}{c|}{\textbf{$\leq$ 0.05\%}} & \multicolumn{2}{c|}{\textbf{0}} \\
        \hline
        \textbf{Option} & \textbf{PV} & \textbf{PV+WT} & \textbf{PV} & \textbf{PV+WT} & \textbf{PV} & \textbf{PV+WT} \\ 
        \hline
        \textbf{NPC (\$)} & 123,412 & 100,416 & 125,432 & 90,542 & 132,616 & 101,530 \\
        \hline
        \textbf{LCOE (\$/kWh)} & 0.711 & 0.579 & 0.723 & 0.522 & 0.764 & 0.585 \\
        \hline
        \textbf{BESS (kWh)} & 27.96 & 18.64 & 37.28 & 18.64 & 37.28 & 18.64 \\
        \hline
        \textbf{PV (kW)} & 15.4 & 8.12 & 11.3 & 7.4 & 13 & 10 \\
        \hline
        \textbf{WT (kW)} & - & 8 & - & 6 & - & 6 \\
        \hline
        \textbf{\% Real Unmet} & 0.06\% & 0.04\% & 0.05\% & 0.05\% & 0 & 0 \\
        \hline
    \end{tabular}
    \vspace{-0.5mm}
\end{table}

Table~\ref{tab:sol_auto_lead} presents the results of the battery autonomy–based microgrid design, where the BESS capacity is predetermined according to the expected autonomy hours~\eqref{eq:auto_hour}. Since the HOMER Pro solution can result in approximately 0.05\% unmet load even when a 0\% unmet load constraint is specified, the developed MILP method adopts a less than 0.05\% unmet load criterion to ensure a fair comparison. It can be observed that the hybrid PV and WT cases are more economical than the PV-only cases. The solutions obtained from HOMER Pro and the developed method are generally close, although the latter achieves better results in some instances, particularly for the hybrid cases under 20.9 autonomy hours, with a 6.3\% cost reduction.
\begin{equation}
    \text{Autonomy hour} = \frac{24\times(SOC_{\text{max}} - SOC_{\text{min}}) E_{\text{rated}}^{\text{bess}}}{\text{Daily Load}}
    \label{eq:auto_hour}
\end{equation}

\begin{table}[!t]
    \caption{Planning results considering autonomy (Lead Acid).}
    %\vspace{-2mm}
    % \vspace{-3mm}
    \label{tab:sol_auto_lead}
    \centering
    \scriptsize
    \setlength{\tabcolsep}{3pt} % Reduce column padding
    \renewcommand{\arraystretch}{1.2} % Adjust row height for better vertical centering
    \begin{tabular}{|c|c|c|c|c|c|c|}
        \hline
        \multicolumn{3}{|c|}{\textbf{BESS Autonomy}} & \multicolumn{4}{c|}{\textbf{NPC (\$)}} \\
        \hline
        \multirow{2}{*}{\textbf{Strings}} & \multirow{2}{*}{\textbf{kWh}} & \multirow{2}{*}{\textbf{Hour}} & \multicolumn{2}{c|}{\textbf{HOMER Pro}} & \multicolumn{2}{c|}{\textbf{Developed}} \\
        \cline{4-7}
        & & & \textbf{PV} & \textbf{PV+WT} & \textbf{PV} & \textbf{PV+WT} \\
        \hline
        1 & 9.32 & 6.96 & \multicolumn{4}{c|}{\textit{Infeasible}} \\
        \hline
        3 & 27.96 & 20.9 & 123,095 & 100,587 & 125,454 (\textit{+2\%}) & 94,275 (\textit{-6.3\%}) \\
        \hline
        6 & 55.92 & 41.8 & 146,832 & 145,002 & 144,828 (\textit{-1.3\%}) & 143,607 (\textit{-1\%}) \\
        \hline
        7 & 65.24 & 48.7 & 161,510 & 164,091 & 161,287 (\textit{-0.1\%}) & 162,603 (\textit{-0.9\%}) \\
        \hline
        9 & 83.88 & 62.6 & 199,529 & 202,267 & 199,702 (\textit{+0.1\%}) & 201,017 (\textit{-0.6\%}) \\
        \hline
        
    \end{tabular}
    \vspace{-4mm}
\end{table}

\subsection{Lead Acid vs Li-ion LFP}
%The cost of Li-ion LFP BESS has decreased over the past decades, making it a suitable option for microgrid applications. 

Table~\ref{tab:sol_opti_comp_bess} compares the optimal NPC and system configurations of the microgrid when deploying two types of BESS: lead-acid and Li-ion LFP. The results are obtained using the developed method. It can be observed that the Li-ion BESS is a more economical choice, with cost reductions of 5.54\% and 7.48\% for PV-only and hybrid microgrid cases, respectively, when no unmet load is allowed. These cost savings increase to 10.13\% and 60.5\% when allowing less than 0.05\% unmet load. Additionally, the Li-ion cases require either fewer renewable resources or a smaller BESS capacity due to their longer lifespan and higher Depth of Discharge (DoD).

\begin{table}[!t]
    \caption{Optimal planning results: Lead Acid vs. Li-ion LFP.}
     %\vspace{-1mm}
    \label{tab:sol_opti_comp_bess}
    \centering
    \scriptsize
    \setlength{\tabcolsep}{2pt} % slightly reduced column padding
    \renewcommand{\arraystretch}{1.1} % slightly tighter row spacing
    \begin{tabular}{|c|c|c|c|c|c|}
        \hline
        \multicolumn{2}{|c|}{\textbf{Cases}} & \multicolumn{2}{c|}{\textbf{PV}} & \multicolumn{2}{c|}{\textbf{PV+WT}} \\
        \hline
        \textbf{Unmet} & \textbf{Comp.} & \textbf{LA} & \textbf{LFP} & \textbf{LA} & \textbf{LFP} \\
        \hline
        \multirow{4}{*}{0} 
        & NPC (\$) & 132,616 & 125,272 (\textit{-5.54\%}) & 101,530 & 93,940 (\textit{-7.48\%}) \\
        \cline{2-6}
        & BESS (kWh) & 37.28 & 31.2 & 18.64 & 18.72 \\
        \cline{2-6}
        & PV (kW) & 13 & 11.8 & 10 & 8.3 \\
        \cline{2-6}
        & WT (kW) & - & - & 6 & 4 \\
        \hline
        \multirow{4}{*}{0.05\%} 
        & NPC (\$) & 125,432 & 112,728 (\textit{-10.13\%}) & 90,542 & 85,064 (\textit{-6.05\%}) \\
        \cline{2-6}
        & BESS (kWh) & 37.28 & 24.96 & 18.64 & 18.72 \\
        \cline{2-6}
        & PV (kW) & 11.3 & 12.4 & 7.4 & 6.2 \\
        \cline{2-6}
        & WT (kW) & - & - & 6 & 4 \\
        \hline
    \end{tabular}
    \vspace{-4mm}
\end{table}

\section{Conclusions}\label{s:conclusions}

This paper presents a validation of the customized microgrid planning tool,  \name, developed using the MILP algorithm, with HOMER Pro used as a benchmark for case studies. The tool offers greater flexibility and accuracy by enabling user-defined constraints and delivering optimal solutions through commercial solvers. Comparative results show that integrating WTs into the hybrid system reduces the NPC by decreasing PV and battery sizing requirements. Sensitivity analysis performed on battery autonomy revealed that as autonomy increases, the cost difference between PV only and hybrid systems narrows, due to the growing dominance of storage costs. Additionally, the comparison between battery technologies showed that LFP batteries are more cost effective than lead acid alternatives in long term applications. 
Overall, \name\ proves to be a financially viable solution for standalone microgrids located in remote and off-grid harsh environments.
Future work will focus on incorporating uncertainty modeling, verifying performance under disturbances, and extending the analysis to diversified regions and different scales of microgrids.
%Overall, \name\ proves to be a reliable and financially viable solution for standalone microgrids located in remote and off-grid harsh environments.

\section*{{Acknowledgments}}
This study was supported by King Abdullah University of Science and Technology (KAUST) and by Saudi Aramco under grant number RGC/3/6504-01-01. 

\ifCLASSOPTIONcaptionsoff
  \newpage
\fi

% \begin{IEEEbiography}{Xxx~Yyy}
% \end{IEEEbiography}

%\bibliographystyle{bibliography/IEEEtranIES}
%\bibliography{bibliography/IEEEabrv,References}
\bibliographystyle{IEEEtran}
\bibliography{References}  % <-- this should match your .bib filename (without .bib extension)

\end{document}